\documentclass[12pt]{iopart}

\usepackage{cite}
\usepackage{mathptmx}
\expandafter\let\csname equation*\endcsname\relax
  \expandafter\let\csname endequation*\endcsname\relax

\usepackage{amsmath}
\usepackage{graphicx}
\usepackage[version=3]{mhchem}

\usepackage{float}
\usepackage{fancyhdr}
\usepackage{fnpos}
\usepackage[english]{babel}

\usepackage{array}
\usepackage{droidsans}
\usepackage{charter}
\usepackage[T1]{fontenc}

\usepackage{setspace}
\usepackage[final]{hyperref} 
\hypersetup{
	colorlinks=true,       
	linkcolor=blue,        
	citecolor=blue,        
	filecolor=magenta,     
	urlcolor=blue         
}

\begin{document}

\title{Interacting Brownian particles exhibiting enhanced rectification in an asymmetric channel}

\author{Narender Khatri and P. S. Burada}

\address{Department of Physics, Indian Institute of Technology Kharagpur, Kharagpur - 721302, India}
\ead{psburada@phy.iitkgp.ac.in}
\vspace{10pt}
\begin{indented}
\item[] July 2021
\end{indented}

\begin{abstract} 
Rectification of interacting Brownian particles is investigated in a two-dimensional asymmetric channel in the presence of an external periodic driving force.
The periodic driving force can break the thermodynamic equilibrium and induces rectification of particles (or finite average velocity).
The spatial variation in the shape of the channel leads to entropic barriers, which indeed control the rectification of particles. 
We find that by simply tunning the driving frequency, driving amplitude, and shape of the asymmetric channel, the average velocity can be reversed.   
Moreover, a short range interaction force between the particles 
further enhances the rectification of particles greatly. 
This interaction force is modeled as the lubrication interaction.
Interestingly, it is observed that there exists a characteristic critical frequency $\Omega_c$ below which the rectification of particles greatly enhances in the positive direction with increasing the interaction strength; whereas, for the frequency above this critical value, it greatly enhances in the negative direction with increasing the interaction strength.
Further, there exists an optimal value of the asymmetric parameter of the channel for which the rectification of interacting particles is maximum.
These findings are useful in sorting out the particles and understanding the diffusive behavior of small particles or molecules in microfluidic channels, membrane pores, etc.
\end{abstract}

\section{Introduction}

In general, in biological systems, solid-state nanopores, and artificial microstructures, it is well known that the Brownian particles move in constrained geometries \cite{Karger, Hille}. 
The diffusive behavior of particles in constrained geometries has not only important theoretical significance but also has potential applications in many processes, especially catalysis, osmosis, particles selectivity, and particles separation \cite{Karger, Hille}.
Note that when the Brownian particles diffuse in a constrained geometry, the spatial confinement to the particles produces an entropic barrier, which significantly impacts the transport properties of the particles \cite{Burada_prl, DR_pre, Khatri_JCP, Khatri_pre, Ai_JSM5, Ai_JSM6, Ai_JSM7, Ai_JSM8, Ao_EPJST_2014, Ghosh_PRL_2013, Li_PRE_2014}. 
In particular, the important examples of constrained geometries where entropic effects are ubiquitous are, for instance, biological cells, ion channels, nanoporous materials, zeolites, microfluidic devices, ratchets, and artificial channels \cite{Hanggi1, Karger, Hille, Han, Matthias}. 

Rectification of Brownian particles in ratchet systems in the presence of an unbiased external force has been an intense area of research in the last few decades \cite{Astumian_science,   Matthias, Hanggi1, Ai, Ai2, Malgaretti_jcp, Ai_JSM1, Ghosh_JSM2, Ai_JSM3, Ai_JSM4, Ai_JSM9}.
Here, the external oscillatory (unbiased) force can induce directed motion of the diffusing particles, i.e., known as rectification, 
in ratchet systems possessing dynamical or spatial symmetry breaking at submicron scales \cite{Hanggi1}.
Particularly, the rectification of Brownian particles in ratchet systems has been inspired by the working principles of Brownian motors \cite{Nori_AnnP}, molecular motors \cite{RD}, nanoscale friction \cite{Krim}, surface smoothing \cite{Derenyi}, coupled Josephson junctions \cite{Hanggi3}, separation of particles \cite{DR}, optical ratchets and directed motion of laser-cooled atoms \cite{Faucheux}, and mass separation and trapping schemes at the microscale \cite{Marchesoni_prl}.
Note that the rectification depends on the physical properties of the diffusing particles such as mass, size, shape, etc. \cite{Hanggi1, Nori_AnnP}.
In order to observe rectification of particles, the ratchet setup requires two basic ingredients \cite{Hanggi4};
(i) asymmetry (spatial and/or temporal) in the system itself 
to violate the left/right symmetry in the diffusive behavior of the particles, and 
(ii) a zero-mean external oscillatory force to break the thermodynamic 
equilibrium, which forbids the appearance of directed motion due to the second law of thermodynamics. 
Depending on various nonequilibrium situations, several models have been proposed to obtain rectification of particles such as rocking ratchets \cite{Magnasco}, flashing ratchets \cite{Hanggi5}, diffusion ratchets \cite{Hanggi6}, and correlation ratchets \cite{Doering}. 
In all these models, the unbiased driving force can break the thermodynamic equilibrium and induces rectification of particles. 
The rectification of non-interacting Brownian particles is a well studied problem both experimentally \cite{Matthias, Hanggi1} and theoretically \cite{Ai, Ai2, Malgaretti_jcp}.
However, the interaction between the particles may further influence the diffusive behavior of particles in confined geometries \cite{Khatri_JCP}.
Usually, the short range interaction force between the Brownian particles 
can be considered as 
(i) the lubrication interaction \cite{Khatri_JCP, Neill}, 
(ii) the harmonic interaction \cite{Ai_H}, 
(iii) the Lennard-Jones interaction \cite{Rein_LJ}, and 
(iv) the Yukawa interaction \cite{Mazroui_Yukawa}.

In this article, we numerically study how the interaction between particles and the channel shape influences 
the rectification of the Brownian particles when they diffuse in an asymmetric periodic channel in the presence of a periodic driving force. 
Since we are dealing with the diffusive behavior of a large number of Brownian particles in a confined environment, out of the above mentioned interactions, 
we consider the lubrication interaction in the current study. 
The lubrication interaction, as the name implies, results from the thin layer of viscous fluid that separates the surfaces of nearly touching Brownian particles \cite{Li_JCP_2020}. In other words, it provides the hydrodynamic effect, originated by the action of the fluid stress, on a particle due to its neighbors. 
Also, it plays a vital role, particularly when the particles are confined. 
Rest of this article is organized as follows. 
In Sec.~\ref{Model}, we introduce the model for the interacting Brownian particles in a two-dimensional asymmetric channel in the presence of 
an external periodic driving force. 
The influences of the particles interaction and the channel shape on rectification are discussed in Sec.~\ref{sec:interaction}.
Finally, the main conclusions are presented in Sec.~\ref{Conclusions}.

\section{Model}\label{Model}

\begin{figure}[htb]
\centering
\includegraphics[scale = 1]{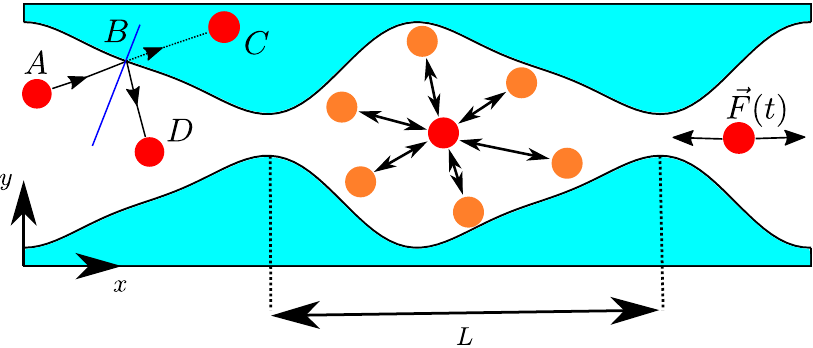}
\caption{Schematic illustration of a two-dimensional asymmetric channel, with the periodicity $L$, confining the motion of Brownian particles which are subjected to an external non-adiabatic oscillatory force $\vec{F}(t) = F_0 \sin(\omega t) \hat{x}$ along the $x$ direction and considering the short range interaction force between the particles.
The reflecting channel boundaries are defined by equation~(\ref{wallf}) which assure the confinement of particles inside the channel.}
\label{fig:chw}
\end{figure}

Consider the motion of a Brownian particle, suspended in a two-dimensional asymmetric channel, driven by an external oscillatory force, and the interaction force $\vec{F}_\mathrm{int}$ due to its neighboring particles (see Fig.~\ref{fig:chw}). 
In the overdamped regime, the equation of motion of the Brownian particle is given by the Langevin equation, 
\begin{equation}\label{Langevin1}
\eta \frac{d \vec{r}}{d t} =  F_0 \sin(\omega t) \, \hat{x} + \vec{F}_\mathrm{int}  + \sqrt{\eta k_B T} \, \vec{\xi}(t), 
\end{equation}
where $\vec{r}$ is the position of the particle in two dimensions, 
$F_0$ is the amplitude of the periodic driving force acting along the $x$ direction only, $\omega$ is the frequency of the driving force, $\eta$ is the friction coefficient, $k_B$ is the Boltzmann constant, and $T$ is the temperature. 
The thermal fluctuations due to the coupling of the particle with the surrounding heat bath are modeled by a zero-mean Gaussian white noise $\vec{\xi}(t)$, obeying the fluctuation-dissipation relation $\langle \xi_i (t)  \xi_j (t') \rangle = 2 \delta_{ij} \delta(t-t')$ for $i, j = x, y$.

We have considered a few types of interaction between the particles, and out of all, we find that the lubrication interaction plays a vital role, particularly when the particles are confined in a channel which is the case in the current study (details not shown). 
Therefore, we consider the lubrication interaction to study the rectification of particles. 
The short range interaction force on a particle $i$ due to its neighbors reads \cite{Khatri_JCP, Neill},  
\begin{equation}\label{Lubrication}
\displaystyle{ \vec{F}_\mathrm{int} =  \sum_{ j = 1, ~j \neq i}^{n} \frac{\sigma_{ij}}{d_{ij}}\left( \cos\theta_{ij}\, \hat{x} + \sin\theta_{ij} \, \hat{y} \right)}, 
\end{equation}
where the sum is taken over all its nearest neighbors, $\sigma_{ij}$ is the interaction strength between the particle $i$ and $j$, $d_{ij}$ is the corresponding distance between the particles, $\theta_{ij}$ is the angle that $d_{ij}$ makes with the channel axis ($x$ axis), and $\hat{x}$ and $\hat{y}$ are the unit vectors along $x$ and $y$ directions, respectively. 
In the limit of high density of identical particles whose size is much smaller than both the periodicity and width of the channel, this interaction force can be approximated as \cite{Khatri_JCP} 
\begin{equation}
\vec{F}_\mathrm{int} = K \, ( \cos \theta \,\hat{x} + \sin \theta \,\hat{y}),
\label{eq:Vicsek}
\end{equation}
where $K$ denotes the total strength of the interaction force on a particle due to its neighbors and $\theta$ is a random variable that can have values between $0$ to $2 \pi$. In this limit, equation~(\ref{eq:Vicsek}) is similar to the Vicsek interaction \cite{Vicsek, Vicsek2}, which has been used extensively to study the collective behavior of active particles. 
In particular, the same model for the lubrication interaction has been previously used to study the biased transport of interacting Brownian particles in symmetric channels \cite{Khatri_JCP}.

The shape of two-dimensional asymmetric and spatially periodic channel is described by its half-width (see Fig.~\ref{fig:chw}) 
\begin{equation} 
\label{wallf}
H(x)= a\left[\sin\left(\frac{2\pi x}{L}\right) + \frac{\Delta }{4} \sin\left(\frac{4\pi x}{L}\right) \right]  + b,  
\end{equation}  
where $L$ corresponds to the periodicity of the channel, $\Delta$ denotes the asymmetric parameter of the channel, and parameters $a$ and $b$ control the slope and channel width at the bottleneck, respectively.
Note that the channel is biased towards the positive $x$ direction for $\Delta > 0$ and negative $x$ direction for $\Delta < 0$.

In order to have a dimensionless description, we henceforth scale all lengths by the periodicity of the channel $L$ and time by $\tau = \eta L^2/(k_B T_R)$, which is the characteristic diffusion time at an arbitrary but irrelevant reference temperature $T_R$ \cite{Burada_ESR_prl}.
In dimensionless form, the Langevin equation reads
\begin{equation}\label{Langevin2}
\frac{d \vec{r}}{d t} = f_0 \sin (\Omega t) \hat{x} + \vec{f}_\mathrm{int}  + \sqrt{ D} \, \vec{\xi}(t). 
\end{equation} 
Here, the noise intensity (or the rescaled temperature) is given by $D = T/T_R$. 
The dimensionless interaction force becomes $ \vec{f}_\mathrm{int} =  \vec{F}_\mathrm{int} L/(k_B T_R)$. 
The other parameters read $f_0 = F_0 L/(k_B T_R)$, $\Omega = \omega \tau$, and $k = K L/(k_B T_R)$. 

While solving the Langevin equation $(\ref{Langevin2})$ numerically, we have considered that the particle elastically reflects at the channel boundary to ensure the confinement within the channel. 
	Figure~\ref{fig:chw} shows the reflection of a particle at the channel boundary. 
	Let us say the initial position of the particle was $A(x_1, y_1)$, and in the next instant of time, the position of particle is at $C(x_2, y_2)$, which is outside of the channel boundary [see Fig.~\ref{fig:chw}]. 
	The line joining points $A$ and $C$ intersects the channel boundary at a point $B(p, q)$, i.e., the reflection point which can be calculated numerically using the bisection method.
	The desired position $D(x_3, y_3)$ of the particle after reflection at point $B$ is given by \cite{Khatri_JCP} 
	\begin{subequations}
		\begin{alignat}{2}
		x_3 & = p \pm \, \frac{l}{\sqrt{1 + m_3^2}},\\
		y_3 & = q + m_3(x_3 - p),
		\end{alignat}
		\label{eq:fxy}
	\end{subequations} 
	where $m_3$ is the slope of line $BD$ and $l = \sqrt{(y_2 - q)^2 + (x_2 - p)^2}$.
	Note that depending on the driving force acting on the particle, the particle may reflect multiple times at the channel boundaries to reach the final position.

Knowingly, the Fokker-Planck equation corresponding to the Langevin equation~(\ref{Langevin2}) cannot be solved analytically for the considered system. 
Therefore, we stick to the Brownian dynamics simulations performed within the stochastic Euler algorithm by the integration of the Langevin equation~(\ref{Langevin2}) using reflecting boundary conditions at the channel walls. 
As initial conditions, we have assumed that at $t = 0$ all the particles are randomly distributed in a cell of the channel.  
In the long time limit, the average velocity of particles, which is an observable of the rectification of particles, along the $x$ direction is given by 
\begin{equation}\label{velocity}
v = \lim_{t\to\infty} \frac{\langle x(t) \rangle}{t}.
\end{equation}
In particular, the positive and negative average velocity means that the particles drift to the positive and negative $x$ directions, respectively. 
Note that the effective diffusion coefficient of particles in the long time limit, which is an observable of the dispersion of particles, along the $x$ direction is calculated as \cite{Burada_prl}
\begin{equation}\label{diffusion}
	D_{eff} = \lim_{t\to\infty} \frac{\langle x^2(t) \rangle - \langle x(t) \rangle ^2}{2 t}.
\end{equation}

\section{Impact of particle-particle interaction}
\label{sec:interaction}

\subsection{Enhanced rectification}
\label{sec:rectification}

\begin{figure}[htb]
\centering
\includegraphics[scale = 0.85]{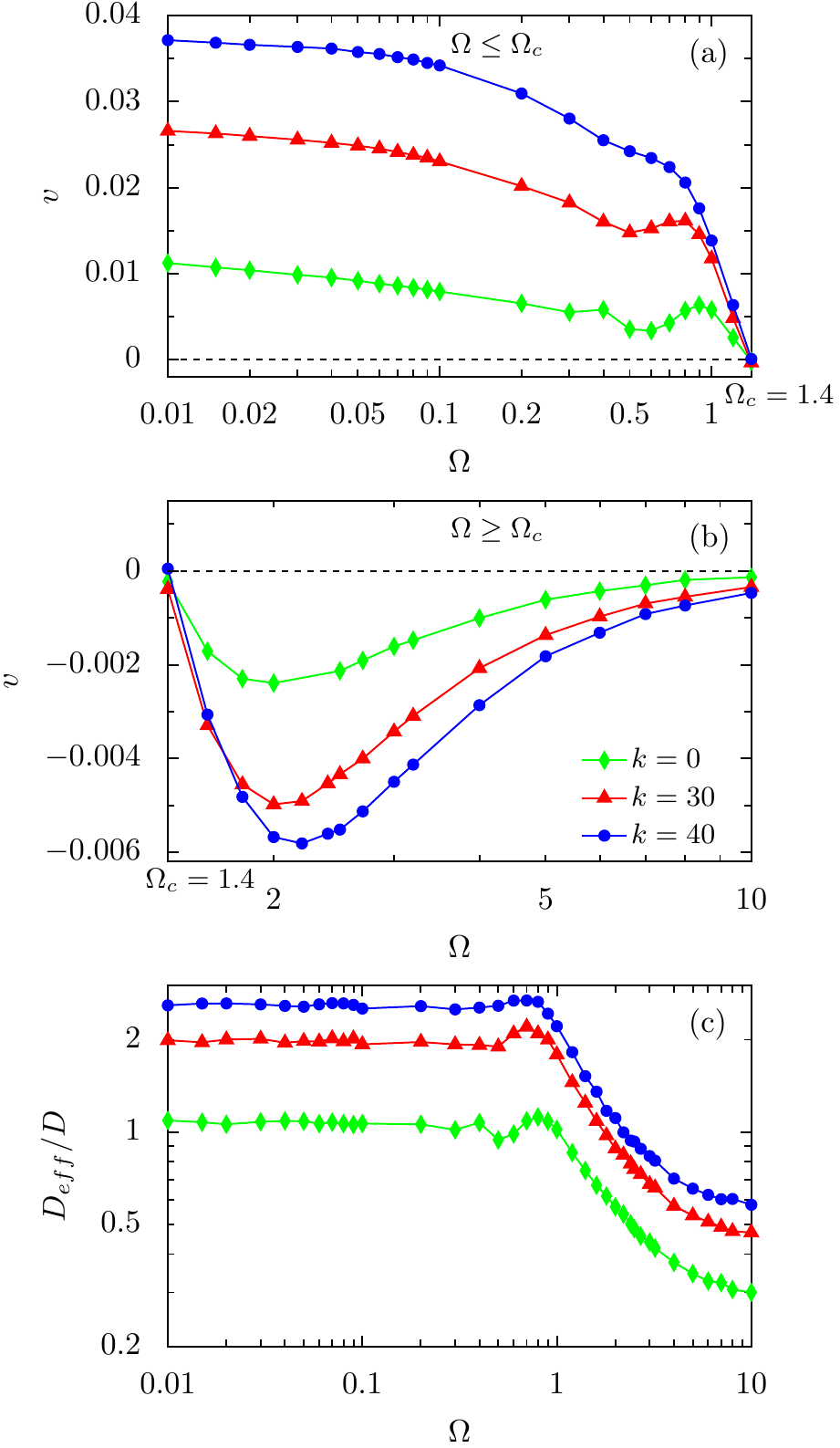}
\caption{Average velocity $v$ versus driving frequency $\Omega$ for various values of the interaction strength between the particles. The value of characteristic critical frequency $\Omega_c$ is approximately 1.4. (a) For $\Omega \leq \Omega_c \approx 1.4$ and (b) for $\Omega \geq \Omega_c \approx 1.4$. 
The corresponding scaled effective diffusion coefficient $D_{eff}/D$ is depicted in (c).	
The set parameters are $a = 1/(2\pi)$, $b = 1.2/(2\pi)$, $\Delta = 1$, $L = 2\pi$, $f_0 = 2.5$, and $D = 0.3$.}
\label{fig:graph2}
\end{figure}

Figures~\ref{fig:graph2}(a)-(b) depict the average velocity $v$ as a function of the driving frequency $\Omega$ for various values of the interaction strength between the particles. 
It is important to point out that there exists a characteristic critical frequency $\Omega_c \approx 1.4$ for which the average velocity is zero.
For $\Omega < \Omega_c$, because of the chosen structure of the channel, 
particles move towards the long end (or the slanted side) resulting in a positive average velocity.
Whereas, for $\Omega > \Omega_c$, due to the higher frequency, 
the particles do not get enough time to cross the long end.
Thus, the particles drift to the short end (or the steeper side) resulting in a negative average velocity.
 Note that the characteristic critical frequency $\Omega_c$ mainly depends on the periodicity of the channel $L$, noise intensity $D$, and amplitude of the oscillating force $f_0$. The same is reported by Ai \cite{Ai_JCP_2009} in the case of non-interacting Brownian particles exhibiting rectification in an asymmetric channel.
Here, we observe that the scaled effective diffusion coefficient $D_{eff}/D$ enhances linearly with increasing the interaction strength $k$ (see Fig.~\ref{fig:graph2}(c)).
This enhancement in $D_{eff}/D$ helps to clarify that dispersion of particles, which facilitates the ratchet transport, increases with increasing $k$.
Therefore, the average velocity greatly enhances in both the positive and negative directions with increasing the interaction strength (see (a) and (b) in Fig.~\ref{fig:graph2}). 
In particular, in the very high frequency limit, i.e., $\Omega \rightarrow \infty$, the average velocity tends to zero irrespective of the interaction strength. 
It is because, in this limit, the particles experience a time-averaged constant force $f = \int_{0}^{\frac{2 \pi}{\Omega}} f(t) dt = 0$.

\begin{figure}[htb!]
	\centering	
	\includegraphics[scale = 0.85]{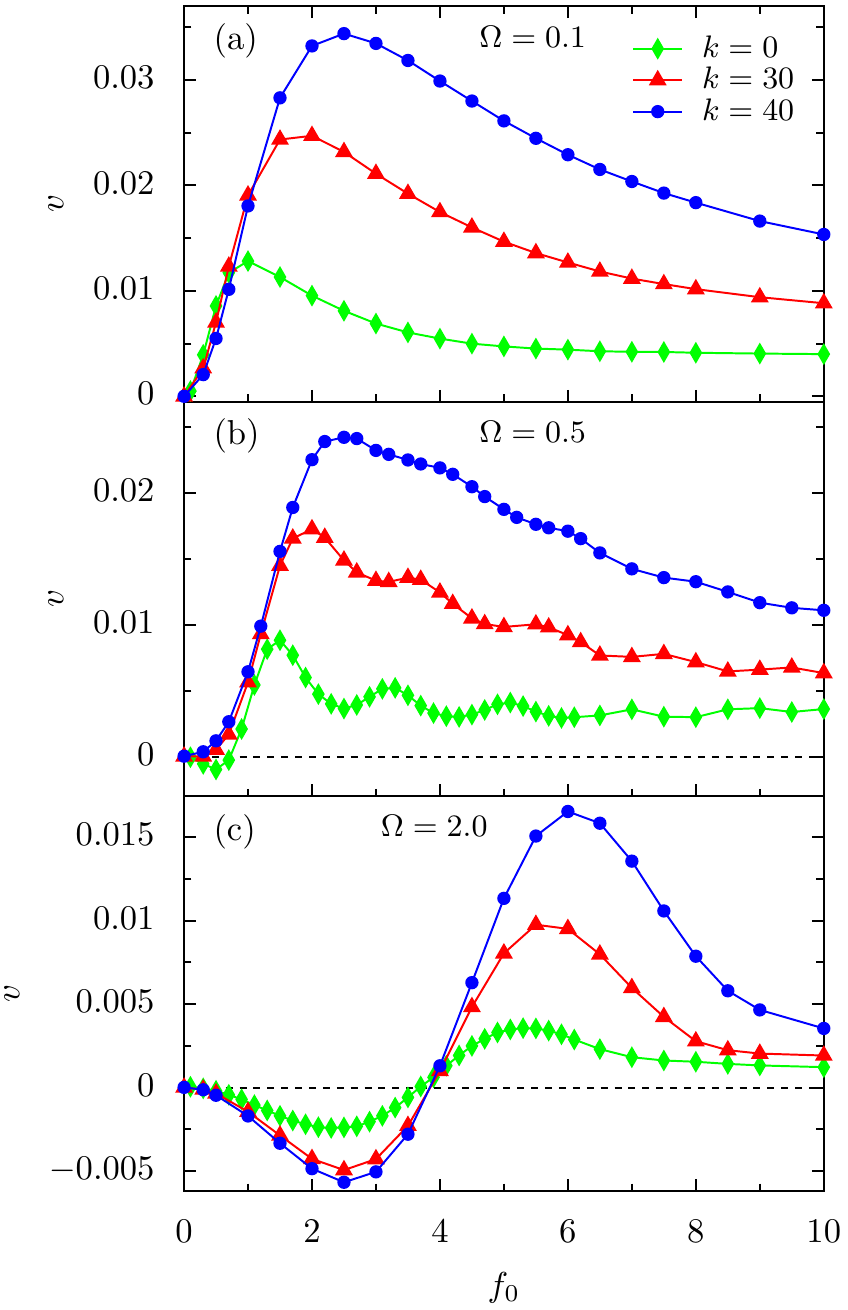}
	\caption{Average velocity $v$ versus driving amplitude $f_0$ for various values of the interaction strength between the particles. (a) For $\Omega = 0.1$, (b) for $\Omega = 0.5$, and (c) for $\Omega = 2.0$. The set parameters are $a = 1/(2\pi$), $b = 1.2/(2\pi)$, $\Delta = 1$, $L = 2\pi$, and $D = 0.3$.}
	\label{fig:graph3}
\end{figure}

Figure~\ref{fig:graph3} depicts the average velocity $v$ as a function of the driving amplitude $f_0$ for various values of the interaction strength $k$. 
For the non-interacting particles, $v$ remains positive as a function of $f_0$ for the lower values of $\Omega$ \cite{Ai2}.
With increasing $\Omega$, $v$ reversal occurs as a function of $f_0$.
As mentioned earlier, the interaction between the particles facilitates the rectification of particles. 
Thus, for $\Omega < \Omega_c$, we observe that $v$ greatly enhances in the positive direction with increasing $k$. 
Whereas, for $\Omega > \Omega_c$, $v$ greatly enhances in both the positive and negative directions with increasing $k$.
Interestingly, $v$ exhibits fluctuations for the moderate $\Omega$ values, e.g., $ \Omega = 0.5$. 
Note that these fluctuations start to disappear with increasing $k$ (see Fig.~\ref{fig:graph3}(b)). 
The latter is due to an increase in the dispersion of particles with $k$.
In particular, for the small and large $f_0$ values, the average velocity tends to zero for various interaction strength values. 
It is because the driving force can be neglected for small driving amplitude values and the effect of the asymmetry of the channel disappears for the large driving amplitude values \cite{Schmid_Adv}.

\begin{figure}[htb!]
\centering	
\includegraphics[scale = 0.85]{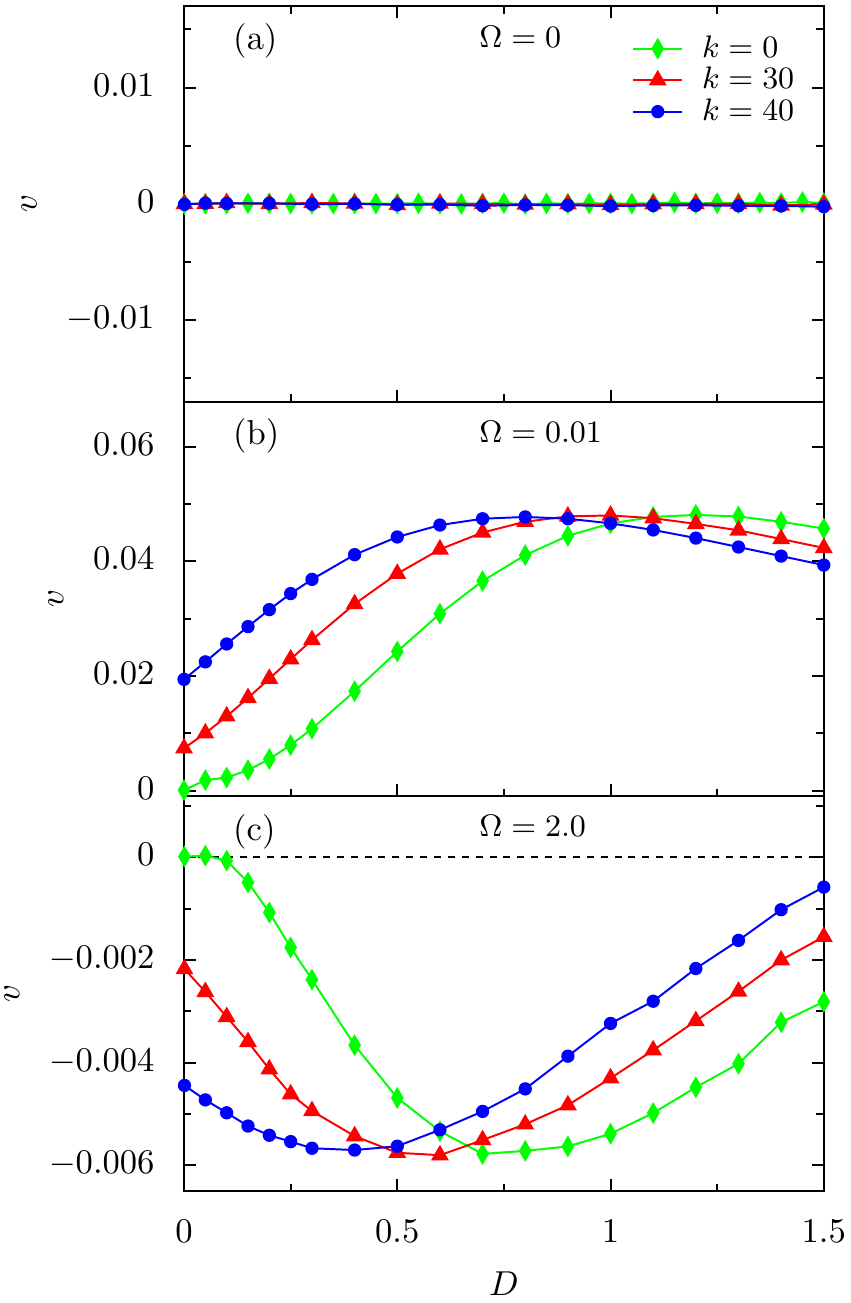}
\caption{Average velocity $v$ versus noise intensity $D$ for various values of the interaction strength between the particles. (a) For $\Omega = 0$, (b) for $\Omega = 0.01$, and (c) for $\Omega = 2.0$. 
The set parameters are $a = 1/(2\pi)$, $b = 1.2/(2\pi)$, $\Delta = 1$, $L = 2\pi$, and $f_0 = 2.5$.}
\label{fig:graph4}
\end{figure}

\begin{figure}[htb!]
	\centering
	\includegraphics[scale = 0.85]{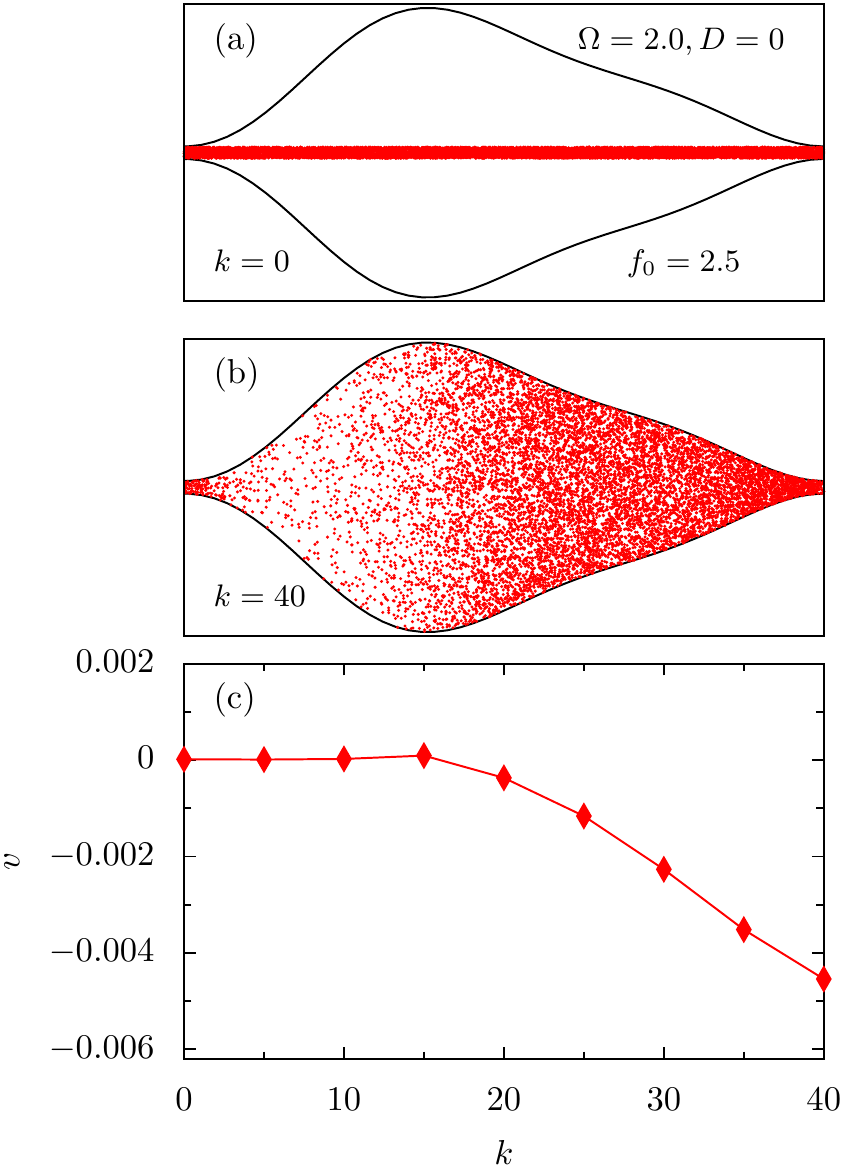}
	\caption{Steady state distribution of the particles mapped into a single cell of the two-dimensional asymmetric channel for two different values of the interaction strength $k$ is depicted in (a) and (b).
		The average velocity $v$ as a function of the interaction strength $k$  is depicted in (c).  
		The set parameters are $a = 1/(2\pi)$, $b = 1.2/(2\pi)$, $\Delta = 1$, $L = 2\pi$, $f_0 = 2.5$, $\Omega = 2.0$, and $D = 0$.}
	\label{fig:graph5}
\end{figure}

Figure~\ref{fig:graph4} depicts the average velocity $v$ as a function of the noise intensity $D$ for various values of the interaction strength $k$ and at different values of the driving frequency $\Omega$. 
For $\Omega = 0$, as one would expect, $v$ is zero for any value of $D$ and $k$ because the driving force is absent.
For the non-interacting particles, $v$ remains positive as a function of $D$ for the lower value of $\Omega$. 
On the other hand, for the higher value of $\Omega$, $v$ remains negative as a function of $D$ (see Fig.~\ref{fig:graph4}(c)). 
When $D = 0$, the non-interacting particles cannot explore the available area in the channel as they move only along the channel direction (see Fig.~\ref{fig:graph5}(a)), and the effect of the asymmetry of the channel disappears, resulting in a zero average velocity.
However, due to the interaction, particles can explore the two-dimensional area and would feel the asymmetry of the channel.
As a result, they exhibit a finite $v$, which increases with $k$. 
On increasing $D$, $v$ greatly enhances in the positive direction with increasing $k$ for $\Omega < \Omega_c$. 
Whereas, for $\Omega > \Omega_c$,  the same is observed in the negative direction. 
Note that there exists a maximum in $v$ at an optimal value of $D$, and it shifts toward the lower values of $D$ with increasing $k$. 
For the higher values of $D$, the interaction effects start disappearing, and $v$ decreases.

To understand the significance of $k$ better, we look at the steady state distribution of particles mapped into a single cell of the two-dimensional asymmetric channel for different values of $k$ when $D = 0$ (see Fig.~\ref{fig:graph5}).
As mentioned before, in the limit $k \to 0$, $v$ is zero.
On further increasing $k$, the particles start exploring the available area in the cell due to the interaction force from their neighbors, and the asymmetry of the channel begins to show its signature.
As mentioned earlier, particles do not get enough time to cross the long end for the higher frequency; hence, most of them enter into the cell from the short end.
As a result, most of the particles are found towards the long end. 
Accordingly, $v$ becomes negative and monotonically increases with increasing $k$ (see Fig.~\ref{fig:graph5}(c)).

\subsection{Impact of asymmetric parameter $\Delta$}
\label{sec:Delta_comparison} 

\begin{figure}[H]
	\centering
	\includegraphics[scale=0.9]{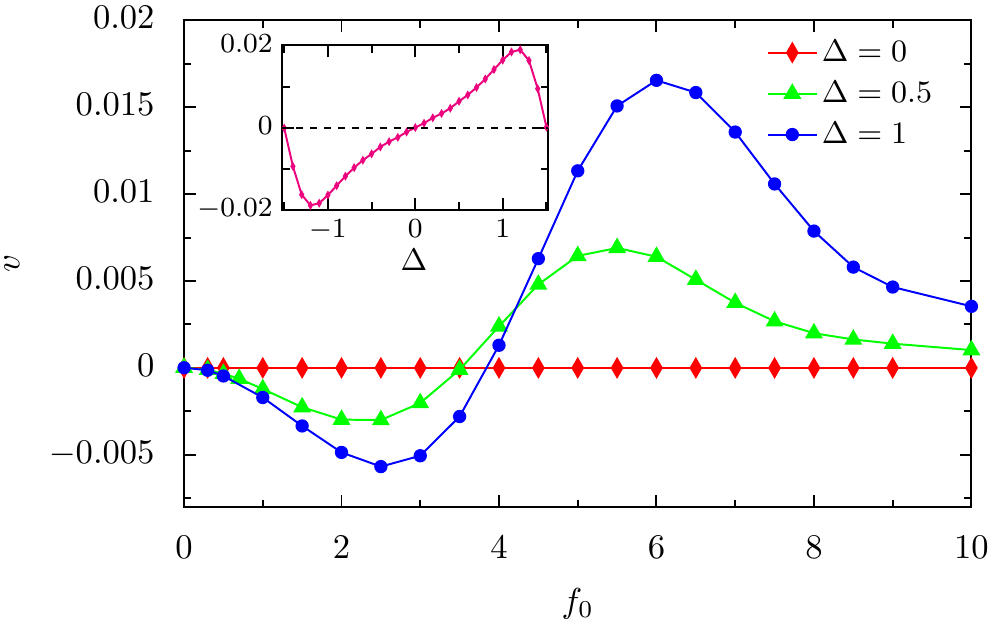}
	\caption{Average velocity $v$ versus driving amplitude $f_0$ for various values of the asymmetric parameter $\Delta$. The inset depicts the dependence of $v$ on $\Delta$ for $f_0 = 6$.   
		The set parameters are $a = 1/(2\pi)$, $b = 1.2/(2\pi)$, $L = 2\pi$, $k = 40$, $D = 0.3$, and $\Omega = 2.0$.}
	\label{fig:graph6}
\end{figure}

Here, we study the impact of the asymmetric parameter $\Delta$ on the rectification of interacting particles. 
In particular, $\Delta$ can have values between -1.5 to 1.5 because for $|\Delta| > 1.5$, the channel is blocked, i.e., the minimum width of the channel becomes zero (see Fig.~\ref{fig:chw}). 
Figure~\ref{fig:graph6} depicts the average velocity $v$ as a function of the driving amplitude $f_0$ for different values of the asymmetric parameter $\Delta$. 
The qualitative behavior of the average velocity as a function of driving amplitude remains the same for $\Delta \ne 0$.
The peculiar behavior of the average velocity as a function of the asymmetric parameter is shown in the inset of 
Figure~\ref{fig:graph6}. 
Note that $\Delta$ controls the direction of the average velocity. 
The average velocity is positive for $\Delta > 0$, zero at $\Delta = 0$, and negative for $\Delta < 0$. 
When $\Delta = 0$, the channel is symmetric, so the average velocity is zero. 
Whereas, when $|\Delta| > 1.5$, the channel is blocked, so the average velocity becomes zero. 
Therefore, there exists an optimal value of $|\Delta|$ for which the average velocity is maximum.

Note that the observed interaction effects can also be studied experimentally by creating a corrugated asymmetric channel by microprinting on a substrate and measuring the drift velocity of the interacting particles \cite{Mahmud_nature}.
These particles can be fabricated from luminescent polystyrene with well defined identical diameters \cite{Matthias}.
In order to have an estimate in real units, which is very useful for the experimentalists, the characteristic values of the main parameters for the overdamped interacting Brownian particles in water moving in a corrugated asymmetric channel with asymmetric parameter $\Delta = 1$ and period length $L \sim 1 ~\mu \mathrm{m}$, and the temperature set at room temperature ($T \sim 300 ~ \mathrm{K}$) are given by $\eta \sim 2 \times 10^{-9} ~ \mathrm{kg/s}$ \cite{Cussler}, $\tau \sim 0.15~ \mathrm{s}$, $k_B T_R/L \sim 10^{-14}~\mathrm{N}$, and $L/\tau \sim 6.67~ \mu \mathrm{m/s}$.
For the parameters $F_0 \sim 2.5 \times 10^{-14} ~\mathrm{N}$, $\omega \sim 0.07 ~ \mathrm{s^{-1}}$, and $ K \sim 4 \times 10^{-13} ~\mathrm{N}$, the interacting particles drift in the positive direction with an average velocity $V \sim 0.25 ~ \mu \mathrm{m/s}$, which is approximately $3.1$ times higher with respect to no interaction case.   
The characteristic critical frequency in the real unit is given by $\omega_c \sim 9.33 ~\mathrm{s^{-1}}$ below which the interacting particles greatly enhance the average velocity in the positive direction, whereas above this critical value it is greatly enhanced in the negative direction.
It is expected that the enhanced rectification can be used for the efficient and selective continuous separation of particles mixture based on their physical properties like mass, size, shape, etc. \cite{Hanggi1,DR,Mukhopadhyay_prl}.
Also, it may have high relevance in describing transport in mesoscopic systems \cite{Astumian_science, Prost_RMP} and controlled drug release \cite{Siegel}.

\section{Conclusions}\label{Conclusions}

In this work, we have numerically studied the rectification of interacting Brownian particles in a two-dimensional asymmetric channel.
With respect to no interaction case, which has been studied earlier both experimentally \cite{Matthias,Hanggi1} and theoretically \cite{Ai,Ai2,Malgaretti_jcp}, we have shown that for the interacting Brownian particles, it is possible to greatly enhance the rectification of particles.
We found that the interaction between particles greatly enhances the average velocity in both the positive and negative directions for various parameter values. 
It has been observed that there exists a characteristic critical frequency $\Omega_c$ below which the average velocity greatly enhances in the positive direction with increasing the interaction strength; however, for the frequency above this critical value, the average velocity greatly enhances in the negative  direction with increasing the interaction strength. 
Moreover, we have shown that there exists an optimal value of the asymmetric parameter of the channel for which the rectification of interacting particles is maximum.
These results have a wide application in many processes, such as separation or spatial sorting of particles mixture based on their physical properties \cite{Hanggi1,DR,Mukhopadhyay_prl}, diffusion of ions and macromolecular solutes through the channels in biological membranes \cite{Hille}, controlled drug release \cite{Siegel}, and many more.
In the future, the current study can be extended to non-Brownian particles under the same situation. 
Particularly, this will be relevant for biological systems where it is quite common to observe anomalous diffusion \cite{Metzler_PCCP}.

\section{Acknowledge}

This work was supported by the Indian Institute of Technology
Kharagpur under the Grant No. IIT/SRIC/PHY/TAB/2015-16/114.

\section*{References}
\bibliography{rsc} 
\bibliographystyle{unsrt} 
\end{document}